\documentclass[12pt, onecolumn]{IEEEtran}
\renewcommand{\baselinestretch}{1.65}

\usepackage{graphicx}

\newtheorem{theorem}{\bf Theorem}
\newtheorem{corollary}{\bf Corollary}

\begin{document}
\title{Fountain Communication using Concatenated Codes}
\author{Zheng Wang, \IEEEmembership{Student Member, IEEE}, Jie Luo, \IEEEmembership{Member, IEEE}
\thanks{The authors are with the Electrical and Computer Engineering Department, Colorado State University, Fort Collins, CO 80523. E-mail: \{zhwang, rockey\}@engr.colostate.edu. }
\thanks{This work was supported by the National Science Foundation under Grant CCF-0728826. Part of the material in this paper were presented at the IEEE International Symposium on Information Theory, Seoul, Korea, June 2009 \cite{ref Wang09}.}
}

\maketitle

\begin{abstract}
This paper extends linear-complexity concatenated coding schemes to fountain communication over the discrete-time memoryless channel. Achievable fountain error exponents for one-level and multi-level concatenated fountain codes are derived. It is also shown that concatenated coding schemes possess interesting properties in several multi-user fountain communication scenarios.
\end{abstract}

%\begin{keywords}
%coding complexity, concatenated codes, error exponent, fountain communication
%\end{keywords}

% Will be ignored for other modes.
\IEEEpeerreviewmaketitle

%\newpage

\section{Introduction}\label{SectionI}
Fountain communication \cite{ref Byers98} is a new communication model proposed for reliable data transmission over channels with arbitrary erasures. In a point-to-point fountain communication system, the transmitter maps a message into an infinite sequence of channel symbols, which experience {\it arbitrary} erasures during transmission. The receiver decodes the message after the number of received symbols exceeds certain threshold. With the help of randomized coding, fountain communication achieves the same rate and error performance over different channel erasure realizations corresponding to an identical number of received symbols. Under the assumption that the erasure statistics is unknown at the transmitter, communication duration in a fountain system is determined by the receiver, rather than by the transmitter.

The first realization of fountain codes was LT codes introduced by Luby \cite{ref Luby02} for erasure channels. LT codes can recover $k$ information bits from $k+O\left(\sqrt{k}\ln^2(k/\delta)\right)$ encoded symbols with probability $1-\delta$ and a complexity of $O(k\ln(k/\delta))$, for any $\delta>0$ \cite{ref Luby02}. Shokrollahi proposed Raptor codes in \cite{ref Shok06} by combining appropriate LT codes with a pre-code. Raptor codes can recover $k$ information bits from $k(1+\epsilon)$ encoded symbols at high probability with complexity $O\left(k\log(1/\epsilon)\right)$. LT codes and Raptor codes can achieve optimum rate with close to linear and linear complexity, respectively. However, under a fixed rate, error probabilities of the two coding schemes do not decrease exponentially in the number of received symbols. Generalization of Raptor codes from erasure channels to binary symmetric channels (BSCs) was studied by Etesami and Shokrollahi in \cite{ref Etesami06}. In \cite{ref Shamai07}, Shamai, Telatar and Verd\'{u} systematically extended fountain communication to arbitrary channels and showed that fountain capacity \cite{ref Shamai07} and Shannon capacity take the same value for stationary memoryless channels. Achievability of fountain capacity was demonstrated in \cite{ref Shamai07} using a random coding scheme whose error probability decreases exponentially in the number of received symbols. Unfortunately, the random coding scheme considered in \cite{ref Shamai07} is impractical due to its exponential complexity.

In classical point-to-point communication over a discrete-time memoryless channel, it is well known that Shannon capacity can be achieved with an exponential error probability scaling law and a linear encoding/decoding complexity \cite{ref Guruswami05}\cite{ref Zheng09}. The fact that communication error probability can decrease exponentially in the codeword length at any information rate below the capacity was firstly shown by Feinstein \cite{ref Feinstein55}. The corresponding exponent was defined as the error exponent. Tight lower and upper bounds on error exponent were obtained by Gallager \cite{ref Gallager65}, and by Shannon, Gallager, Berlekamp \cite{ref Shannon67}, respectively. In \cite{ref Forney66}, Forney proposed a one-level concatenated coding scheme that combines a Hamming-sense error correction outer code with Shannon-sense random inner channel codes. One-level concatenated codes can achieve a positive error exponent, known as the Forney's exponent, for any rate less than Shannon capacity with a polynomial complexity \cite{ref Forney66}. Forney's concatenated codes were generalized by Blokh and Zyablov \cite{ref Blokh82} to multi-level concatenated codes, whose maximum achievable error exponent is known as the Blokh-Zyablov error exponent. In \cite{ref Guruswami05}, Guruswami and Indyk introduced a class of linear complexity near maximum distance separable (MDS) error-correction codes. By using Guruswami-Indyk's codes as the outer codes in concatenated coding schemes, achievability of Forney's and Blokh-Zyablov exponents with linear coding complexity over general discrete-time memoryless channels was proved in \cite{ref Zheng09}.

In this paper, we show that classical concatenated coding schemes can be extended to fountain communication over the discrete-time memoryless channel to achieve positive fountain error exponent (defined in Section \ref{SectionII}) at any rate below the fountain capacity with a linear coding complexity. Achievable error exponents for one-level and multi-level concatenated fountain codes are derived. We show that these error exponents are close in value to their upper bounds, which are Forney's exponent \cite{ref Forney66} for one-level concatenation and Blokh-Zyablov exponent \cite{ref Blokh82} for multi-level concatenation, respectively. We also show that concatenated fountain codes possess several interesting properties useful for network applications. More specifically, when one or more transmitters send common information to multiple receivers over discrete-time memoryless channels, concatenated fountain codes can often achieve near optimal rate and error performance simultaneously for all receivers even if the receivers have different prior knowledge about the transmitted message.

The rest of the paper is organized as follows. The fountain communication model is defined in Section \ref{SectionII}. In Section \ref{SectionIII}, we introduce the preliminary results on random fountain codes, which are basic components of the concatenated coding schemes. One-level and multi-level concatenated fountain codes are introduced in Section \ref{SectionIV}. Special properties of concatenated fountain codes in network communication scenarios are introduced in Sections \ref{SectionV} and \ref{SectionVI}. The conclusions are given in Section \ref{SectionVII}. We use natural logarithms throughout this paper.

\section{Fountain Communication Model}
\label{SectionII}

\begin{figure}[h]
  \centering
  \includegraphics[width=5 in]{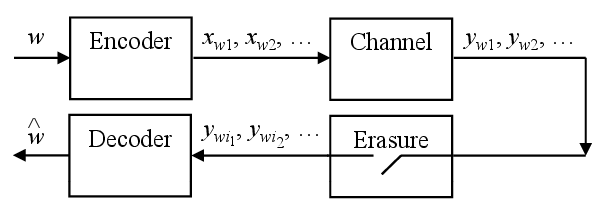}
  \caption{\label{Fig1} Fountain communication over a memoryless channel.}
\end{figure}

Consider the fountain communication system illustrated in Figure \ref{Fig1}.
%Figure 1 goes here.
Assume that the encoder uses a fountain coding scheme \cite{ref Shamai07} with $W$ codewords to map the source message $w\in \{1,2, \cdots, W\}$ into an infinite channel input symbol sequence $\{x_{w1}, x_{w2}, \cdots \}$. Assume that the channel is discrete-time memoryless, characterized by the conditional point mass function (PMF) or probability density function (PDF) $p_{Y|X}(y|x)$, where $x\in \mathcal X$ is the channel input symbol with $\mathcal X$ being the finite channel input alphabet, and $y\in \mathcal Y$ is channel output symbol with $\mathcal Y$ being the finite channel output alphabet, respectively. Assume that the channel information is known at both the encoder and the decoder\footnote{The case when channel information is not available at the encoder will be investigated in Section \ref{SectionVI}.}. The channel output symbols are then passed through an erasure device which generates arbitrary erasures. Define schedule $\mathcal{N}=\{i_1, i_2, \cdots, i_{|{\cal N}|}\}$ as a subset of positive integers, where $|\mathcal{N}|$ is its cardinality \cite{ref Shamai07}. Assume that the erasure device generates erasures only at those time instances not belonging to schedule $\mathcal N$. In other words, only the channel output symbols with indices in $\mathcal N$, denoted by $\{ y_{wi_1}, y_{wi_2}, \cdots, y_{wi_{|{\cal N}|}}\}$, are observed by the receiver. The schedule $\mathcal N$ is arbitrarily chosen and unknown at the encoder.

Rate and error performance variables of the system are defined as follows. We say the fountain rate of the system is $R = (\log W)/N$, if the decoder, after observing $|{\cal N}|= N$ channel symbols, outputs an estimate $\hat{w} \in \{1,2,\cdots,W\}$ of the source message based on $\{ y_{wi_1}, y_{wi_2}, \cdots, y_{wi_{|{\cal N}|}}\}$ and ${\cal N}$. Decoding error happens when $\hat{w}\ne w$. Define error probability $P_e(N)$ as,
\begin{equation}\label{ErrorProb}
P_e(N)=\max_{w}\sup_{\mathcal{N},|\mathcal{N}|\ge N}Pr\{\hat{w}\neq w|w,\mathcal{N}\}.
\end{equation}
We say a fountain rate $R$ is {\it achievable} if there exists a fountain coding scheme with $\lim_{N\rightarrow\infty}P_e(N)=0$ at rate $R$ \cite{ref Shamai07}. The exponential rate at which error probability vanishes is defined as the fountain error exponent, denoted by $E_F(R)$,
\begin{equation}
E_F(R)=\lim_{N\to \infty} -\frac{1}{N}\log P_e(N).
\end{equation}
Define fountain capacity $\mathcal{C}_F$ as the supremum of all achievable fountain rates. It was shown in \cite{ref Shamai07} that $\mathcal{C}_F$ equals Shannon capacity of the stationary memoryless channel. Note that the scaling law here is defined with respect to the number of {\em received} symbols.

\section{Random Fountain Codes}
\label{SectionIII}

In a random fountain coding scheme \cite{ref Shamai07}, encoder and decoder share a fountain code library $\mathcal{L}=\{C_\theta : \theta \in \Theta\}$, which is a collection of fountain codebooks $C_\theta$ indexed by a set $\Theta$. All codebooks in the library have the same number of codewords and each codeword has an infinite number of channel input symbols. Let $C_{\theta}(w)_j$ be the $j^{th}$ codeword symbol in codebook $C_{\theta}$ corresponding to message $w$, for $j\in\{1,2,\cdots\}$. To encode the message, the encoder first selects a codebook by generating $\theta$ according to a distribution $\vartheta$, such that the random variables $x_{w,j}:\theta\rightarrow C_{\theta}(w)_j$ are i.i.d. with a pre-determined input distribution $p_X$ \cite{ref Shamai07}. Then the encoder uses codebook $C_{\theta}$ to map the message into a codeword. We assume that the actual realization of $\theta$ is known to the decoder but is unknown to the erasure device. Therefore channel erasures, although arbitrary, are independent from the codebook generation. Maximum likelihood decoding is assumed at the decoder given the knowledge of the codebook, schedule, and channel information \cite{ref Shamai07}. Due to the random codebook selection, without being conditioned on $\theta$, the error probability experienced by each message is identical. Therefore, the error probability $P_e(N)$ defined in (\ref{ErrorProb}) can be written as follows \cite{ref Shamai07},
\begin{equation}
P_e(N)=\max_{w}\sup_{\mathcal{N},|\mathcal{N}|\ge N}Pr\{\hat{w}\neq w|w,\mathcal{N}\}= \sup_{\mathcal{N},|\mathcal{N}|\ge N}\frac{1}{W}\sum_{w}Pr\{\hat{w}\neq w|w,\mathcal{N}\}.
\label{AverageErrorProb}
\end{equation}

\begin{theorem}\label{Theorem1} Consider fountain communication over a discrete-time memoryless channel $p_{Y|X}$. Let $\mathcal{C}_F$ be the fountain capacity. For any fountain rate $R<\mathcal{C}_F$, random fountain codes achieve the following random-coding fountain error exponent
\begin{equation}
E_{Fr}(R)=\max_{p_X}E_{FL}(R, p_X),
\label{EFr}
\end{equation}
where $E_{FL}(R, p_X)$ is defined as
\begin{eqnarray}
&&E_{FL}(R, p_X)=\max_{0\le \rho \le 1} \left\{-\rho R+ E_0(\rho, p_X)\right\}, \nonumber \\
&&E_0(\rho, p_X)=-\log\sum_y \left( \sum_x p_X(x)p_{Y|X}(y|x)^{\frac{1}{1+\rho}}\right)^{(1+\rho)}.
\label{EFL}
\end{eqnarray}
If the channel is continuous, then summations in (\ref{EFL}) should be replaced by integrals. $\QED$
\end{theorem}

Theorem \ref{Theorem1} was claimed implicitly in, and can be shown by, the proof of \cite[Theorem 2]{ref Shamai07}.

$E_{Fr}(R)$ given in (\ref{EFr}) equals the random-coding exponent of a classical communication system over the same channel \cite{ref Gallager65}. For binary symmetric channels (BSCs), since random linear codes simultaneously achieve the random-coding exponent at high rates and the expurgated exponent at low rates \cite{ref Barg02}, it can be easily shown that the same fountain error exponent is achievable by random linear fountain codes. However, it is not clear whether there exists an expurgation operation, such as the one proposed in \cite{ref Gallager65}, that is robust to the observation of any subset of channel outputs. Therefore, whether the expurgated exponent is achievable for fountain communication over a general discrete-time memoryless channel is unknown.

\section{Concatenated Fountain Codes}
\label{SectionIV}

\begin{figure}[h]
  \centering
  \includegraphics[width=5 in]{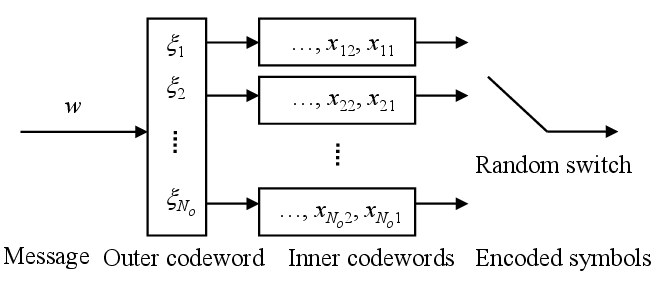}\caption{\label{Fig2} One-level concatenated fountain codes.}
\end{figure}

Consider a one-level concatenated fountain coding scheme illustrated in Figure \ref{Fig2}.
%Figure 2 goes here
Assume that source message $w$ can take $\lfloor\exp(NR)\rfloor$ possible values with equiprobability, where $R$ is the targeted fountain information rate. Assume that the communication terminates after $N$ channel output symbols are observed at the decoder. The one-level concatenated fountain code consists of an outer code and several inner codes. The encoder first encodes the message using the outer code into an outer codeword $\{\xi_1, \xi_2, \cdots, \xi_{N_o}\}$, with $N_o$ outer symbols, each belonging to a finite field of appropriate size. We assume that the outer code is a linear-time encodable/decodable near MDS error-correction code of rate $r_o \in (0,1]$. That is, at a fixed $r_o$ and as $N_o$ is taken to infinity, the outer code can recover the source message from a received codeword with $dN_o$ symbol erasures and $tN_o$ symbol errors, so long as $2t+d \le (1-r_o-\zeta_0)$, where $\zeta_0>0$ is a positive constant that can be made arbitrarily small. The encoding and decoding complexities are linear in the number of outer codeword length $N_o$. An example of such linear complexity error-correction code was presented by Guruswami and Indyk in \cite{ref Guruswami05}. Each outer symbol $\xi_k$ $(k\in\{1,\cdots, N_o\})$ can take $\left\lfloor\exp\left(\frac{N}{N_o}\frac{R}{r_o}\right)\right\rfloor$ possible values.

We use a set of random fountain codes described in Section \ref{SectionIII} as the inner codes, each with $\lfloor\exp(N_iR_i)\rfloor$ codewords, where $N_i=\frac{N}{N_o}$ and $R_i=\frac{R}{r_o}$. To simplify the notations, we have assumed that $N_i$ and $N_o$ are both integers. We also assume that $N_o \gg N_i \gg 1$. The encoder then uses these inner codes to map each outer symbol $\xi_k$ into an inner codeword, which is an infinite sequence of channel input symbols $\{x_{k1}, x_{k2}, \cdots\}$. The inner codewords are regarded as $N_o$ channel input symbol queues, as shown in Figure \ref{Fig2}. In each time unit, the encoder uses a random switch to pick one inner code and sends the first channel input symbol in the corresponding queue through the channel as modeled in Section \ref{SectionII}. The transmitted symbol is then removed from the queue. We use $\theta$ to index the realization of the compounded randomness of codebook generation and switch selection. Let $C^{(k)}_{\theta}(\xi_k)_j$ be the $j^{th}$ codeword symbol of the $k^{th}$ inner code in codebook $\mathcal C_{\theta}^{(k)}$, corresponding to $\xi_k$. Let $Z_{l,\theta}\in\{1,\cdots,N_o\}$ be index of the queue that the random switch chooses at the $l^{th}$ time unit for $l\in\{1,2,\cdots\}$. We assume that index $\theta$ is generated according to a distribution $\vartheta$ such that random variables $x_{k, \xi_k,j}:\theta \rightarrow C^{(k)}_{\theta}(\xi_k)_j$ are i.i.d. with a pre-determined input distribution $p_X$, random variables $I_l:\theta \rightarrow Z_{l,\theta}$ are i.i.d. uniform, $x_{k, \xi_k,j}$ and $I_l$ are independent. The decoder is assumed to know the outer codebook and the code libraries of the inner codes. We also assume that the decoder knows the exact codebook used for each inner code and the exact order in which channel input symbols are transmitted.

Decoding starts after $N=N_oN_i$ channel output symbols are received. The decoder first distributes the received symbols to the corresponding inner codes. Assume that, for $k\in\{1,\cdots,N_o\}$, $z_kN_i$ channel output symbols are received from the $k^{th}$ inner code, where $z_k>0$ and $z_kN_i$ is an integer. We term $z_k$ the ``effective codeword length parameter" of the $k^{th}$ inner code. By definition, we have $\sum_{k=1}^{N_o}z_k=N_o$. Based on $z_k$, and the received channel output symbols, $\{y_{ki_1}, y_{ki_2}, \dots, y_{ki_{z_kN_i}}\}$, the decoder computes the maximum likelihood estimate $\hat{\xi}_k$ of the outer symbol $\xi_k$ together with an optimized reliability weight $\alpha_k\in [0, 1]$. We assume that, given $z_k$ and  $\{y_{ki_1}, y_{ki_2}, \cdots, y_{ki_{z_kN_i}}\}$, reliability weight $\alpha_k$ is computed using Forney's algorithm presented in \cite[Section 4.2]{ref Forney66}. With $\{\hat{\xi_k}\}$ and $\{\alpha_k\}$ for all $k$, the decoder then carries out a generalized minimum distance (GMD) decoding of the outer code and outputs an estimate $\hat{w}$ of the source message. GMD decoding of the outer code here is the same as that in a classical communication system, the detail of which can be found in \cite{ref Zheng09}.

Due to random codebook selection and random switching, without being conditioned on $\theta$, error probabilities experienced by all messages are equal, i.e., $P_e(N)$ satisfies (\ref{AverageErrorProb}). Compared with a classical concatenated code where all inner codes have the same length, in a concatenated fountain coding scheme, numbers of received symbols from different inner codes may be different. Consequently, error exponent achievable by one-level concatenated fountain codes, given in the following theorem, is less than Forney's exponent.

\begin{theorem}\label{Theorem2}
Consider fountain communication over a discrete-time memoryless channel $p_{Y|X}$ with fountain capacity $\mathcal{C}_F$. For any fountain rate $R<\mathcal{C}_F$, the following fountain error exponent can be arbitrarily approached by one-level concatenated fountain codes,
\begin{equation}
E_{Fc}(R)=\max_{p_X, \frac{R}{\mathcal{C}_F}\le r_o \le 1, 0\le \rho \le 1}(1-r_o)\left( -\rho \frac{R}{r_o} + E_0(\rho, p_X)\left[1 -\frac{1+r_o}{2}E_0(\rho, p_X)\right] \right),
\label{EFC}
\end{equation}
where $E_0(\rho, p_X )$ is defined in (\ref{EFL}).

Encoding and decoding complexities of the one-level concatenated codes are linear in the number of transmitted symbols and the number of received symbols, respectively. $\QED$
\end{theorem}

The proof of Theorem \ref{Theorem2} is given in Appendix \ref{ProofTheorem2}.

\begin{corollary}\label{Corollary1}
$E_{Fc}(R)$ is upper-bounded by Forney's error exponent $E_c(R)$ given in \cite{ref Forney66}, and is lower-bounded by $\tilde{E}_{Fc}(R)$, defined by
\begin{equation}
\tilde{E}_{Fc}(R)=\max_{p_X, \frac{R}{\mathcal{C}_F}\le r_o \le 1, 0\le \rho \le 1}(1-r_o)\left( -\rho \frac{R}{r_o} + E_0(\rho, p_X)\left[1 -E_0(\rho, p_X)\right] \right).
\label{TildeEFC}
\end{equation}
The bounds are asymptotically tight in the sense that $\lim_{R\to \mathcal{C}_F} \frac{\tilde{E}_{Fc}(R)}{E_{Fc}(R)} =1$. $\QED$
\end{corollary}

The proof of Corollary \ref{Corollary1} is given in Appendix \ref{ProofCorollary1}.

In Figure \ref{Fig3}, we illustrate $E_{Fc}(R)$, $E_c(R)$, and  $\tilde{E}_{Fc}(R)$ for a BSC with crossover probability $0.1$.
%Figure 3 goes here
We can see that $E_{Fc}(R)$ is closely approximated by $\tilde{E}_{Fc}(R)$, especially at rates close to the fountain capacity.

Extending the one-level concatenated fountain codes to the multi-level concatenated fountain codes is essentially the same as in classical communication systems \cite{ref Blokh82}\cite{ref Zheng09} except that random fountain codes are used as inner codes in a fountain system. For a positive integer $m$, the achievable error exponent of an $m$-level concatenated fountain codes is given in the following Theorem.

\begin{theorem}\label{Theorem3}
Consider fountain communication over a discrete-time memoryless channel $p_{Y|X}$ with fountain capacity $\mathcal{C}_F$. For any fountain rate $R<\mathcal{C}_F$, the following fountain error exponent can be arbitrarily approached by an $m$-level $(m\in \{1,2,\cdots\})$ concatenated fountain codes,
\begin{eqnarray}
&& E_{Fc}^{(m)}(R)=\max_{p_X, \frac{R}{C_F}\le r_o \le 1} \frac{\frac{R}{r_o}-R}{\frac{R}{r_om}\sum_{i=1}^m \left[E_{FL}\left(\left(\frac{i}{m}\right)\frac{R}{r_o}, p_X \right) \right]^{-1}}, \nonumber \\
&& E_{FL}\left(x, p_X \right)=\max_{0\le \rho \le 1} \left( -\rho x + E_0(\rho, p_X)\left[1 -E_0(\rho, p_X)\right] \right),
\label{EFCm}
\end{eqnarray}
where $E_0(\rho, p_X )$ is defined in (\ref{EFL}).

For a given $m$, the encoding and decoding complexities of the $m$-level concatenated codes are linear in the number of transmitted symbols and the number of received symbols, respectively.
$\QED$
\end{theorem}

Theorem \ref{Theorem3} can be proved by following the analysis of $m$-level concatenated codes presented in \cite{ref Blokh82}\cite{ref Barg05} and replacing the error exponent of code in each concatenation level  with the corresponding error exponent lower bound given in Corollary \ref{Corollary1}.

\begin{corollary}\label{Corollary2}
The following fountain error exponent can be arbitrarily approached by multi-level concatenated fountain codes with linear encoding/decoding complexity,
\begin{equation}
E_{Fc}^{(\infty)}(R)=\max_{p_X, \frac{R}{\mathcal{C}_F}\le r_o \le 1} \left(\frac{R}{r_o}-R \right)\left[\int_0^{\frac{R}{r_o}} \frac{dx}{E_{FL}\left(x, p_X \right) } \right]^{-1},
\label{EFCInfty}
\end{equation}
where $E_{FL}\left(x, p_X \right)$ is defined in (\ref{EFCm}). $\QED$
\end{corollary}

In Figure \ref{Fig4}, we illustrate $E_{Fc}^{(\infty)}(R)$ and the Blokh-Zyablov exponent $E_c^{(\infty)}(R)$ for a BSC with crossover probability $0.1$.
%Figure 4 goes here
It can be seen that $E_{Fc}^{(\infty)}(R)$ does not deviate significantly from the Blokh-Zyablov exponent, which is the error exponent upper bound for multi-level concatenated fountain codes.

\section{Rate Compatible Fountain Communication}\label{SectionV}
In this section, we consider the fountain communication where the receiver already has partial knowledge about the transmitted message. Take the application of software patch distribution as an example. When a significant number of patches are released, the software company may want to combine the patches together as a service pack. However, if a user already has some of the patches, he may only want to download the new patches, rather than the whole service pack. On one hand, for the convenience of the patch server, all patches of the service pack should be encoded jointly. On the other hand, for the communication efficiency of each particular user, we also want the fountain system to achieve the same rate and error performance as if only the novel part of the service pack is transmitted. We require such performance objective to be achieved simultaneously for all users, and define such a fountain communication model as the rate compatible fountain communication. We will show next that efficient rate compatible fountain communication can be achieved using a class of extended concatenated fountain codes with linear complexity.

Assume that a source message $w$, which takes $\lfloor\exp(NR)\rfloor$ possible values, is partitioned into $L$ sub-messages $[w_1, w_2, \cdots, w_L]$, where $w_i$ $(i\in\{1,\cdots, L\})$ can take $\lfloor\exp(Nr_i)\rfloor$ possible values with $\sum_i r_i =R$. Consider the following extended one-level concatenated fountain coding scheme. For each $i \in \{1, \cdots, L\}$, the encoder first uses a near MDS outer code with length $N_o$ and rate $r_o$ to encode sub-message $w_i$ into an outer codeword $\{\xi_{i1}, \cdots, \xi_{iN_o}\}$, as illustrated in Figure \ref{Fig5}.
%Figure 5 goes here
Next, for all $k \in \{1, \cdots, N_o\}$, the encoder combines outer codeword symbols $\{\xi_{1k}, \cdots, \xi_{L k}\}$ into a macro symbol $\xi_k=[\xi_{1k}, \cdots, \xi_{L k}]$. A random fountain code is then used to map $\xi_k$ into an infinite channel input sequence $\{x_{k1}, x_{k2}, \cdots \}$.

Without loss of generality, we assume that there is only one decoder (receiver) and it already has sub-messages $\{w_{l+1}, \cdots, w_L\}$, where $l \in [1, L-1]$ is an integer. The decoder estimates the source message after $N_l = N\frac{\sum_{i=1}^l r_i}{R}$ channel output symbols are received\footnote{Assume that $N_l$ and $N_l/N_o$ are both integers.}. From the decoder's point of view, since the unknown messages $[w_{1}, \cdots, w_l]$ can only take $\lfloor\exp(N\sum_{i=1}^l r_i)\rfloor$ possible values, the {\it effective} fountain information rate of the system is $R_{ef}=\frac{N\sum_{i=1}^l r_i}{N_l}=R$. According to the known messages $\{w_{l+1}, \cdots, w_L\}$, the decoder first strikes out from fountain codebooks all codewords corresponding to the wrong messages. The extended one-level concatenated fountain code is then decoded using the same procedure as described in Section \ref{SectionIV}. Assume that the average number of symbols received by each inner codeword $\tilde{N}_i=\frac{N_l}{N_o}= \frac{N}{N_o}\frac{\sum_{i=1}^l r_i}{R}$ is large enough to enable asymptotic analysis. By following a similar analysis given in the proof of Theorem \ref{Theorem2}, it can be seen that error exponent $E_{Fc}(R)$ given in (\ref{EFC}) can still be arbitrarily approached.

Therefore, given a rate partitioning $R =[r_1, \cdots, r_L]$, the encoder can encode the complete message irrespective of the sub-messages known at the decoder. The fountain system can achieve the same rate and error performance as if only the unknown sub-messages are encoded and transmitted. If the system has multiple receivers with different priori sub-messages, the rate and error performance tradeoff as characterized in Theorem \ref{Theorem2} can be achieved simultaneously for all receivers. Extending this scheme to the multi-level concatenated codes is straightforward.

\section{Fountain Communication over An Unknown Channel}
\label{SectionVI}
In previous sections, we have assumed that concatenated fountain codes should be optimized based on a known discrete-time memoryless channel model $p_{Y|X}$. However, such an optimization may face various challenges in practical applications. For example, suppose that a transmitter broadcasts encoded symbols to multiple receivers simultaneously. Channels experienced by different receivers may be different. Even if the channels are known, the transmitter still needs to optimize fountain codes simultaneously for multiple channels. For another example, suppose that the source message (e.g., a software patch) is available at multiple servers. A user may collect encoded symbols from multiple servers separately over different channels and use these symbols to jointly decode the message. By regarding the symbols as received over a virtual channel, we want the fountain system to achieve good rate and error performance without requiring the full statistical model of the virtual channel at the transmitter. We term the communication model in the latter example the rate combining fountain communication. In both examples, the research question is whether key coding parameters can be determined without full channel knowledge at the transmitter. In this section, we show that, even when the channel state is unknown at the transmitter, it is still possible to achieve near optimal rate and error performance using concatenated fountain codes.

Consider fountain communication over a discrete-time memoryless channel $p_{Y|X}$ using one-level concatenated fountain codes. We assume that the channel is symmetric, and hence the optimal input distribution $p_X$ is known at the transmitter. Other than channel alphabets and the symmetry property, we assume that channel information $p_{Y|X}$ is unknown at the transmitter, but known at the receiver. Given $p_X$, define $I(p_X)=I(X;Y)$ as the mutual information between the input and output of the memoryless channel. We assume that the transmitter and the receiver agree on achieving a fountain information rate of $\gamma I(p_X)$ where $\gamma \in [0,1]$ is termed the normalized fountain rate, known at the transmitter.

Recall from the proof of Theorem \ref{Theorem2} that, if $p_{Y|X}$ is known at the transmitter, the outer code rate $r_o$ can be predetermined at the transmitter and the following error exponent can be arbitrarily approached,
\begin{eqnarray}
&& E_{Fc}(\gamma, p_X)=\max_{0\le r_o\le 1} E_{Fc}(\gamma, p_X, r_o), \nonumber \\
&& E_{Fc}(\gamma, p_X, r_o)=\max_{0\le \rho\le 1} (1-r_o)I(p_X) \left( -\rho \frac{\gamma}{r_o} + \frac{E_0(\rho, p_X)}{I(p_X)}\left[1 -\frac{1+r_o}{2}E_0(\rho, p_X)\right] \right).
\label{EFCGammapXro}
\end{eqnarray}
Without $p_{Y|X}$ at the transmitter, the optimal $r_o$ cannot be derived. However, with the knowledge of $\gamma$, we can set a suboptimal outer code rate by letting $r_o=\frac{\sqrt{\gamma^2+8\gamma}-\gamma}{2}$ and define the corresponding error exponent by
\begin{eqnarray}
E_{Fcs}(\gamma, p_X)=E_{Fc}\left(\gamma, p_X, r_o=\frac{\sqrt{\gamma^2+8\gamma}-\gamma}{2}\right).
\label{EFCSGammapXro}
\end{eqnarray}
The following theorem indicates that $E_{Fcs}(\gamma, p_X)$ approaches $E_{Fc}(\gamma, p_X)$ asymptotically as $\gamma \to 1$.

\begin{theorem}\label{Theorem4}
Given the discrete-time memoryless channel $p_{Y|X}$ and a source distribution $p_X$, the following limit holds,
\begin{equation}\label{LimitofTheorem4}
\lim_{\gamma \to 1} \frac{E_{Fcs}\left(\gamma, p_X\right)}{E_{Fc}(\gamma, p_X)}=1.
\end{equation}
$\QED$
\end{theorem}

The proof of Theorem \ref{Theorem4} is given in Appendix \ref{ProofTheorem4}.

In Figure \ref{Fig6}, we plot $E_{Fcs}(\gamma, p_X)$ and $E_{Fc}(\gamma, p_X)$ for BSC with crossover probability $0.1$. It can be seen that setting $r_o$ at $r_o=\frac{\sqrt{\gamma^2+8\gamma}-\gamma}{2}$ is near optimal for all normalized fountain rate values.
%Figure 6 goes here
Indeed, computer simulations suggest that such optimality conclusion applies to a wide range of channels over a wide range of fountain rates. However, further investigation on this issue is outside the scope of this paper.

\section{Conclusions}\label{SectionVII}
We extended linear-complexity concatenated codes to fountain communication over a discrete-time memoryless channel. Fountain error exponents achievable by one-level and multi-level concatenated codes were derived. It was shown that the fountain error exponents are less than but close to Forney's and Blokh-Zyablov exponents. In rate compatible communication where decoders know part of the transmitted message, with the encoder still encoding the complete message, concatenated fountain codes can achieve the same rate and error performance as if only the novel part of the message is encoded for each individual user. For one-level concatenated codes and for certain channels, it was also shown that near optimal error exponent can be achieved with an outer code rate independent of the channel statistics.

\appendix
\subsection{Proof of Theorem \ref{Theorem2}}
\label{ProofTheorem2}
\begin{proof}
We first introduce the basic idea of the proof.

Assume that the decoder starts decoding after receiving $N=N_oN_i$ symbols, where $N_o$ is the length of the outer codeword, $N_i$ is the {\it expected} number of received symbols from each inner code. In the following error exponent analysis, we will obtain asymptotic results by first taking $N_o$ to infinity and then taking $N_i$ to infinity.

Let $\mbox{\boldmath $z$}$ be an $N_o$-dimensional vector whose $k^{th}$ element $z_k$ is the effective codeword length parameter of the $k^{th}$ inner code, for $k\in \{1, \cdots, N_o\}$. Note that $\mbox{\boldmath $z$}$ is a random vector. Let $dz>0$ be a small constant. We define $\{z_g|z_g=n dz, n=0,1,\dots, \}$ as the set of ``grid values" each can be written as an non-negative integer multiplying $dz$. Define a point mass function (PMF) $f_Z^{(dz)}$ as follows. We first quantize each element of $\mbox{\boldmath $z$}$, for example $z_k$, to the closest grid value no larger than $z_k$. Denote the quantized $\mbox{\boldmath $z$}$ vector by $\mbox{\boldmath $z$}^{(q)}$, whose elements are denoted by $z^{(q)}_i$ for $i\in \{1, \cdots, N_o\}$. For any grid value $z_g$, we define ${\cal I}_{z_g}=\left\{i\left|z^{(q)}_i=z_g\right.\right\}$ as the set of indices corresponding to which the elements of $\mbox{\boldmath $z$}^{(q)}$ vector equal the particular $z_g$. Given $\mbox{\boldmath $z$}$, the empirical PMF $f_Z^{(dz)}$ is a function defined for the grid values, with $f_Z^{(dz)}(z_g)=\frac{|{\cal I}_{z_g}|}{N_o}$, where $|{\cal I}_{z_g}|$ is the cardinality of ${\cal I}_{z_g}$. Since $f_Z^{(dz)}$ is induced from random vector $\mbox{\boldmath $z$}$, itself is random. Let $Pr\left\{f_Z^{(dz)} \right\}$ denote the probability that the received effective inner codeword length parameter vector $\mbox{\boldmath $z$}$ gives a particular PMF $f_Z^{(dz)}$.

Let us now consider a decoding algorithm, called ``$dz$-decoder", which is the same as the one introduced in Section \ref{SectionIV} except that the decoder, after receiving $N_iz_k$ symbols for the $k^{th}$ inner code (for all $k\in \{1, \cdots, N_o\}$), only uses the first $N_iz_k^{(q)}$ symbols to decode the inner code. Assume that the fountain information rate $R$, the outer code rate $r_o$, and the input distribution $P_X$ are given. Due to symmetry, it is easy to see that, without being conditioned on random variable $\theta$ (defined in Section \ref{SectionIV}), different $\mbox{\boldmath $z$}$ vectors corresponding to the same $f_Z^{(dz)}$ (which is indeed induced from $\mbox{\boldmath $z$}^{(q)}$) give the same error probability performance. Let $P_e\left(f_Z^{(dz)}\right)$ be the communication error probability of the $dz$-decoder given $f_Z^{(dz)}$. Communication error probability $P_e$ of the $dz$-decoder {\it without} given $f_Z^{(dz)}$ can be written as,
\begin{equation}
P_e= \sum_{f_Z^{(dz)}}P_e\left(f_Z^{(dz)}\right)Pr\left\{f_Z^{(dz)}\right\}.
\label{UnconditionalPe}
\end{equation}

For a given $f_Z^{(dz)}$, define $E_{f}(f_{Z}^{(dz)})=-\lim_{N_i\to\infty}\lim_{N_o\to \infty}\frac{1}{N_iN_o}\log P_e\left(f_Z^{(dz)}\right)$. Consequently, we can find a constant $K_0(N_i, N_o)$, such that the following inequality holds for all $f_Z^{(dz)}$ and all $N_i$, $N_o$,
\begin{equation}
P_e\left(f_Z^{(dz)}\right) \le K_0(N_i, N_o) \exp\left(-N_i N_o E_{f}\left(f_{Z}^{(dz)}\right)\right), \qquad \lim_{N_i\to \infty}\lim_{N_o\to \infty}\frac{\log K_0(N_i, N_o)}{N_iN_o}=0.
\label{BoundK0}
\end{equation}

Given $dz$, $N_i$, $N_o$, let $K_1(N_i, N_o)$ be the total number of possible quantized $\mbox{\boldmath $z$}^{(q)}$ vectors (the quantized vector of $\mbox{\boldmath $z$}$). $K_1(N_i, N_o)$ can be upper bounded by
\begin{equation}
K_1(N_i, N_o)\le 2^{N_o} \frac{\left(\left\lceil \frac{N_o}{dz} \right\rceil+N_o-1\right)!}{\left(\left\lceil \frac{N_o}{dz} \right\rceil\right)!(N_o-1)!}.
\label{SubBoundK1}
\end{equation}
In the above bound, the term $\frac{\left(\left\lceil \frac{N_o}{dz} \right\rceil+N_o-1 \right)!}{\left(\left\lceil \frac{N_o}{dz}\right\rceil\right)!(N_o-1)!}$ represents the total number of possible outcomes of assigning $\left\lceil \frac{N_o}{dz} \right\rceil$ identical balls to $N_o$ distinctive boxes. This is the number of possible $\mbox{\boldmath $z$}^{(q)}$ vectors we can get if the received symbols are assigned to the inner codes in groups with $N_i dz$ (assumed to be an integer) symbols per group. Let us term the assumption of assigning received symbols in groups the ``symbol-grouping" assumption. To relax the symbol-grouping assumption, we note that, if the number of symbols obtained by an inner code, say the $k^{th}$ inner code, is a little less that an integer multiplication of $N_i dz$, then the quantization value $z_k^{(q)}$ obtained without the symbol-grouping assumption can be one unit less than the corresponding value with the symbol-grouping assumption. Therefore, the total number of possible $\mbox{\boldmath $z$}^{(q)}$ vectors we can get without the symbol-grouping assumption is upper bounded by $2^{N_o}$ multiplying the corresponding number with the symbol-grouping assumption. Note that, given $dz$, the right hand side of (\ref{SubBoundK1}) is not a function of $N_i$, and it is also an upper bound on the total number of possible $f_Z^{(dz)}$  functions.

Due to Stirling's approximation \cite{ref Romik00}, (\ref{SubBoundK1}) implies that $\lim_{N_o\to \infty}\frac{\log K_1(N_i, N_o)}{N_o} < \infty$, and hence
\begin{equation}
\lim_{N_i\to\infty}\lim_{N_o\to \infty}\frac{\log K_1(N_i, N_o)}{N_iN_o} =0.
\label{BoundK1}
\end{equation}

Combining (\ref{UnconditionalPe}), (\ref{BoundK0}) and (\ref{BoundK1}), the error exponent of a $dz$-decoder is given by
\begin{eqnarray}
E_{Fc}= -\lim_{N_i\to\infty}\lim_{N_o\to \infty}\frac{\log P_e}{N_i N_o}= \min_{f_{Z}^{(dz)}}\left\{E_{f}\left(f_{Z}^{(dz)}\right) - \lim_{N_i\to \infty}\lim_{N_o\to \infty} \frac{1}{N_iN_o} \log  Pr\left\{f_{Z}^{(dz)}\right\} \right \}.
\label{EFCR1}
\end{eqnarray}

The rest of the proof contains four parts. In Part I, the expression of $\lim_{N_i\to \infty}\lim_{N_o\to \infty} \frac{1}{N_iN_o} \log  Pr\left\{f_{Z}^{(dz)}\right\}$ is derived. In Part II, we derive the expression of $E_{f}\left(f_{Z}^{(dz)}\right)$. In Part III, we use the results of the first two parts to obtain $\lim_{dz\to 0}E_{Fc}$. Complexity and the achievable error exponent of the concatenated fountain code is obtained based on the derived results in Part IV.

{\bf Part I: }
Let $\mbox{\boldmath $z$}(i)$ (for all $i\in \{1, \cdots, N_o\}$)  be an $N_o$-dimensional vector with only one non-zero element corresponding to the $i^{th}$ received symbol. If the $i^{th}$ received symbol belongs to the $k^{th}$ inner code, then we let the $k^{th}$ element of $\mbox{\boldmath $z$}(i)$ equal $1$ and let all other elements equal $0$. Since the random switch (illustrated in Figure \ref{Fig2}) picks inner codes uniformly, we have
\begin{equation}\label{Moments}
E[\mbox{\boldmath $z$}(i)]=\frac{1}{N_o}\mbox{\boldmath $1$}, \quad \mbox{cov}[\mbox{\boldmath $z$}(i)]=\frac{1}{N_o}\mbox{\boldmath $I$}_{N_o}-\frac{1}{{N_o}^2}\mbox{\boldmath $1$}\mbox{\boldmath $1$}^T,
\end{equation}
where $\mbox{\boldmath $1$}$ is an $N_o$-dimensional vector with all elements being one, and $\mbox{\boldmath $I$}_{N_o}$ is the identity matrix of size $N_o$. According to the definitions, we have $\mbox{\boldmath $z$}=\frac{1}{N_i}\sum_{i=1}^{N_iN_o}\mbox{\boldmath $z$}(i)$. Since the total number of received symbols equal $N_iN_o$, we must have $\mbox{\boldmath $1$}^T\mbox{\boldmath $z$}=N_o$.

Let $\mbox{\boldmath $\omega$}$ be a real-valued $N_o$-dimensional vector whose entries satisfy $ -\pi\sqrt{N_iN_o}\le \omega_k< \pi\sqrt{N_iN_o}, \forall k\in \{1, \cdots, N_o\} $. Since $\mbox{\boldmath $z$}$ equals the normalized summation of $N_iN_o$ independently distributed vectors $\mbox{\boldmath $z$}(i)$, the characteristic function of $\sqrt{\frac{N_i}{N_o}}(\mbox{\boldmath $z$}-\mbox{\boldmath $1$})$, denoted by $\varphi_Z(\mbox{\boldmath $\omega$})=E\left[\exp\left(j\sqrt{\frac{N_i}{N_o}}\mbox{\boldmath $\omega$}^T(\mbox{\boldmath $z$}-\mbox{\boldmath $1$}) \right)\right]$, can therefore be written as
\begin{eqnarray}
&& \varphi_Z(\mbox{\boldmath$\omega$}) =E\left[\exp\left(j\sqrt{\frac{N_i}{N_o}}\mbox{\boldmath $\omega$}^T(\mbox{\boldmath $z$}-\mbox{\boldmath $1$}) \right)\right]  = \prod_{i=1}^{N_iN_o} E\left[\exp\left(j\sqrt{\frac{1}{N_iN_o}}\mbox{\boldmath $\omega$}^T(\mbox{\boldmath $z$}(i)-\frac{1}{N_o}\mbox{\boldmath $1$}) \right)\right] \nonumber \\
&& = \left\{E\left[\exp\left(j\sqrt{\frac{1}{N_iN_o}}\mbox{\boldmath $\omega$}^T(\mbox{\boldmath $z$}(i)-\frac{1}{N_o}\mbox{\boldmath $1$}) \right)\right]\right\}^{N_iN_o} =\left[1-\frac{1}{2}\frac{\|\mbox{\boldmath $Q$}^T\mbox{\boldmath $\omega$}\|^2}{N_o^2N_i}+ o\left(\frac{\|\mbox{\boldmath $Q$}^T\mbox{\boldmath $\omega$}\|^2}{N_o^2N_i}\right) \right]^{N_oN_i},
\label{CharacteristicFunction}
\end{eqnarray}
where in the last equality, $\mbox{\boldmath $Q$}$ is a real-valued $N_o \times (N_o-1)$-dimensional matrix satisfying $\mbox{\boldmath $Q$}^T\mbox{\boldmath $Q$}=\mbox{\boldmath $I$}_{N_o-1}$ and $\mbox{\boldmath $Q$}^T\mbox{\boldmath $1$}=\mbox{\boldmath $0$}$, which imply $\mbox{\boldmath $Q$}\mbox{\boldmath $Q$}^T=\mbox{\boldmath $I$}_{N_o}-\frac{1}{N_o}\mbox{\boldmath $1$}\mbox{\boldmath $1$}^T $. In other words, $\|\mbox{\boldmath $Q$}^T\mbox{\boldmath $\omega$}\|^2= \mbox{\boldmath $\omega$}^T(\mbox{\boldmath $I$}_{N_o}-\frac{1}{N_o}\mbox{\boldmath $1$}\mbox{\boldmath $1$}^T ) \mbox{\boldmath $\omega$}$.

Note that, since $\mbox{\boldmath $z$}$ is discrete-valued, $\varphi_Z(\mbox{\boldmath$\omega$})$ is similar to a multi-dimensional discrete-time Fourier transform of the PMF of $\sqrt{\frac{N_i}{N_o}}(\mbox{\boldmath $z$}-\mbox{\boldmath $1$})$. Because $\sqrt{N_i N_o}\left[\sqrt{\frac{N_i}{N_o}}(\mbox{\boldmath $z$}-\mbox{\boldmath $1$})\right]=\sum_{i=1}^{N_iN_o}\mbox{\boldmath $z$}(i)-N_i \mbox{\boldmath $1$} $ takes integer-valued entries, the $\varphi_Z(\mbox{\boldmath$\omega$})$ function is periodic $\mbox{\boldmath$\omega$}$ in the sense that $\varphi_Z\left(\mbox{\boldmath$\omega$}+2\pi\sqrt{N_iN_o}\mbox{\boldmath $e$}_{k} \right)=\varphi_Z(\mbox{\boldmath$\omega$})$, $k\in \{1, \cdots, N_o\}$, where $\mbox{\boldmath $e$}_{k}$ is an $N_o$-dimensional vector whose $k^{th}$ entry is one and all other entries are zeros. This is why we can focus on ``frequency" vector $\mbox{\boldmath$\omega$}$ with $ -\pi\sqrt{N_iN_o}\le \omega_k< \pi\sqrt{N_iN_o}, \forall k\in \{1, \cdots, N_o\} $.

Equation (\ref{CharacteristicFunction}) implies that
\begin{equation}\label{LimitMGF}
\lim_{N_o\rightarrow \infty}\left\{\varphi_Z(\mbox{\boldmath$\omega$})-\exp\left(-\frac{1}{2N_o}\mbox{\boldmath$\omega$}^T\mbox{\boldmath $Q$}\mbox{\boldmath $Q$}^T\mbox{\boldmath$\omega$}\right)\right\}=0.
\label{CharacteristicApproximation}
\end{equation}
Therefore, with large enough $N_o$ and for any $\mbox{\boldmath $z$}$, the probability $Pr\{\mbox{\boldmath $z$}\}$ is upper-bounded by
\begin{eqnarray}
Pr\{\mbox{\boldmath $z$}\}&& \le \left(\frac{1}{2\pi\sqrt{N_iN_o}}\right)^{N_o}  \left(\frac{N_o}{2\pi}\right)^{\frac{N_o-1}{2}}\exp\left(-\frac{N_i}{2} \left[\|\mbox{\boldmath $z$}- \mbox{\boldmath $1$}\|^2-dz\|\mbox{\boldmath $1$}\|^2 \right] \right),
\label{DensityY0}
\end{eqnarray}
where the constant $2\pi \sqrt{N_iN_o}$ in the denominator of the first term on the right hand side of (\ref{DensityY0}) is due to the range of $ -\pi\sqrt{N_iN_o}\le \omega_k< \pi\sqrt{N_iN_o}, \forall k\in \{1, \cdots, N_o\} $. The constant $dz\|\mbox{\boldmath $1$}\|^2$ in the exponent of (\ref{DensityY0}) is added to ensure the existence of a large enough $N_o$ to satisfy the inequality, as implied by (\ref{CharacteristicApproximation}). Inequality (\ref{DensityY0}) further implies that
\begin{eqnarray}
Pr\{\mbox{\boldmath $z$}\}&& \le \left(\frac{1}{2\pi\sqrt{N_iN_o}}\right)^{N_o}  \left(\frac{N_o}{2\pi}\right)^{\frac{N_o-1}{2}}\exp\left(-\frac{N_i}{2} \left[\|\mbox{\boldmath $z$}^{(q)}- \mbox{\boldmath $1$}\|^2-3dz\|\mbox{\boldmath $1$}\|^2 \right] \right),
\label{DensityY}
\end{eqnarray}
where $\mbox{\boldmath $z$}^{(q)}$ is the quantized version of $\mbox{\boldmath $z$}$. Consequently, the probability of $\mbox{\boldmath $z$}^{(q)}$ is upper-bounded by
\begin{eqnarray}
Pr\left\{\mbox{\boldmath $z$}^{(q)}\right\}&& \le \left\lceil N_i dz\right\rceil ^{N_o} \left(\frac{1}{2\pi\sqrt{N_iN_o}}\right)^{N_o}  \left(\frac{N_o}{2\pi}\right)^{\frac{N_o-1}{2}}\exp\left(-\frac{N_i}{2} \left[\|\mbox{\boldmath $z$}^{(q)}- \mbox{\boldmath $1$}\|^2-3N_odz \right] \right).
\label{DensityYq}
\end{eqnarray}
The probability of any PMF $f_Z^{(dz)}$ is upper-bounded by
\begin{eqnarray}
&& Pr\left\{f_Z^{(dz)}\right\} \le K_1(N_i, N_o) Pr\left\{\mbox{\boldmath $z$}^{(q)}\right\} \nonumber \\
&& \quad \le  K_1(N_i, N_o)\left\lceil N_i dz\right\rceil ^{N_o} \left(\frac{1}{2\pi\sqrt{N_iN_o}}\right)^{N_o}  \left(\frac{N_o}{2\pi}\right)^{\frac{N_o-1}{2}}\exp\left(-\frac{N_i}{2} \left[\|\mbox{\boldmath $z$}^{(q)}- \mbox{\boldmath $1$}\|^2-3N_odz \right] \right),
\label{DensityY2}
\end{eqnarray}
where $K_1(N_i, N_o)$ is the total number of possible $\mbox{\boldmath $z$}^{(q)}$ vectors satisfying (\ref{BoundK1}).

From (\ref{DensityY2}), we can see that for all $f_Z^{(dz)}$ the following inequality holds,
\begin{equation}
-\lim_{N_i\to \infty}\lim_{N_o\to \infty} \frac{\log Pr\left\{f_Z^{(dz)}\right\}}{N_iN_o} \ge \frac{1}{2}\sum_{z_g} \left[(z_g-1)^2-3dz\right]f_Z^{(dz)}(z_g),
\label{SecondTerm}
\end{equation}
where $f_Z^{(dz)}(z_g)$ is the value of PMF $f_Z^{(dz)}$ at $z_g$.

Note that, because $\mbox{\boldmath $1$}^T\mbox{\boldmath $z$}=N_o$, for all empirical PMFs $f_Z^{(dz)}$, we have $\sum_{z_g} z_g f_Z^{(dz)}(z_g) \in [1-dz, 1]$.

{\bf Part II: }
Next, we will derive the expression of $E_{f}\left(f_Z^{(dz)}\right)$, which is the error exponent conditioned on an empirical PMF $f_Z^{(dz)}$.

Let $\mbox{\boldmath $z$}$ be a particular $N_o$-dimensional effective inner codeword length parameter vector following the empirical PMF $f_Z^{(dz)}$, under a given $dz$. Let $P_e(\mbox{\boldmath $z$})$ be the error probability given $\mbox{\boldmath $z$}$ (or $\mbox{\boldmath $z$}^{(q)}$). Let $P_e(f_Z^{(dz)})$ be the error probability given $f_Z^{(dz)}$. From the definition of the concatenated fountain codes, we can see that the inner codes are logically equivalent, so do the codeword symbols of the near MDS outer code. In other words, error probabilities corresponding to all $\mbox{\boldmath $z$}$ vectors with the same PMF $f_Z^{(dz)}$ are equal. This consequently implies that $P_e(\mbox{\boldmath $z$})=P_e(f_Z^{(dz)})$. Therefore, when bounding $E_{f}\left(f_Z^{(dz)}\right)$, instead of assuming a particular $f_Z^{(dz)}$ which corresponds to multiple $\mbox{\boldmath $z$}$ vectors, we can assume a single $\mbox{\boldmath $z$}$ vector whose corresponding empirical PMF is $f_Z^{(dz)}$.

Assume that the outer code has rate $r_o$, and is able to recover the source message from $dN_o$ outer symbol erasures and $tN_o$ outer symbol errors so long as $d+2t \le (1-r_o -\zeta_0)$, where $\zeta_0 >0$ is a constant satisfying $\lim_{N_i \to \infty}\lim_{N_o \to \infty} \zeta_0 =0$. An example of such near MDS code was introduced in \cite{ref Guruswami05}. Assume that, for all $k$, the $k^{th}$ outer codeword symbol is $\xi_k$, and the $k^{th}$ inner code reports an estimate of the outer symbol $\hat{\xi}_k$ together with a reliability weight $\alpha_k \in [0, 1]$. Applying Forney's GMD decoding to the outer code \cite{ref Zheng09}, the source message can be recovered if the following inequality holds \cite[Theorem 3.1b]{ref Forney66},
\begin{equation}
\sum_{k=1}^{N_o} \alpha_k \mu_k > (r_o+\zeta_0) N_o,
\end{equation}
where $\mu_k=1$ if $\hat{\xi}_k=\xi_k$, and $\mu_k=-1$ if $\hat{\xi}_k \ne \xi_k$. Consequently, error probability conditioned on the given $\mbox{\boldmath $z$}$ vector is bounded by
\begin{eqnarray}
P_e(f_Z^{(dz)})&=&P_e(\mbox{\boldmath $z$}) \le Pr\left\{\sum_{k=1}^{N_o} \alpha_k \mu_k \le (r_o+\zeta_0)N_o \right\} \le \min_{s\ge 0} \frac{E\left[\exp\left(-sN_i\sum_{k=1}^{N_o} \alpha_k \mu_k\right)\right]}{\exp(-sN_i(r_o+\zeta_0)N_o)},
\label{PeRroz1}
\end{eqnarray}
where the last inequality is due to Chernoff's bound.

Given the effective inner codeword length parameter vector $\mbox{\boldmath $z$}$, random variables $\alpha_k \mu_k$ for different inner codes are independent. Therefore, (\ref{PeRroz1}) can be further written as
\begin{eqnarray}
P_e(f_Z^{(dz)})&=&P_e(\mbox{\boldmath $z$}) \le \min_{s\ge 0} \frac{\prod_{k=1}^{N_o} E\left[\exp\left(-sN_i \alpha_k \mu_k\right)\right]}{\exp(-sN_i(r_o+\zeta_0)N_o)} = \min_{s\ge 0} \frac{\exp\left( \sum_{k=1}^{N_o} \log E\left[\exp\left(-sN_i \alpha_k \mu_k\right)\right]\right)}{\exp(-sN_i(r_o+\zeta_0)N_o)}.
\label{PeRroz2}
\end{eqnarray}

Now we will derive the expression of $\log E\left[\exp\left(-sN_i \alpha_k \mu_k\right)\right]$ for the $k^{th}$ inner code.

Assume that the effective codeword length parameter is $z_k$. Given $z_k$, whose quantized value is $z_k^{(q)}$, depending on the received channel symbols, the decoder generates the maximum likelihood outer code estimate $\hat{\xi}_k$, and generates $\alpha_k$ using Forney's algorithm presented in \cite[Section 4.2]{ref Forney66}. Define an adjusted error exponent function $E_z(z)$ as follows.
\begin{eqnarray}
&& E_z(z)=\max_{0\le \rho\le 1} -\rho\frac{R}{r_o} + z E_0(\rho, p_X),
\label{Ez}
\end{eqnarray}
where $E_0(\rho, p_X)$ is defined in (\ref{EFL}). By following Forney's error exponent analysis presented in \cite[Section 4.2]{ref Forney66}, we obtain
\begin{eqnarray}
-\log E\left[\exp\left(-sN_i \alpha_k \mu_k\right)\right]  \ge \max\left\{\min \{N_iE_z\left(z_k^{(q)}\right), N_i\left(2E_z\left(z_k^{(q)}\right)-s\right), N_is \}, 0 \right\} + K_2(N_i, N_o),
\label{EzZk}
\end{eqnarray}
where $K_2(N_i, N_o)$ is a constant satisfying $\lim_{N_i\to \infty}\lim_{N_o\to \infty} \frac{K_2(N_i, N_o)}{N_iN_o}=0$.

Define a function $\phi(z, s)$ as follows,
\begin{equation}
\phi(z, s)=\left\{\begin{array}{ll} -s r_o & z, E_z(z) < s/2 \\ 2E_z(z)-(1+r_o)s & z, s/2\le E_z(z) < s \\ (1-r_o)s & z, E_z(z) \ge s \end{array} \right. .
\end{equation}
Substitute (\ref{EzZk}) into (\ref{PeRroz2}), and take $N_i$, $N_o$ to infinity (which implies $\zeta_0\to 0$), we get the following bound on the conditional error exponent $E_{f}\left(f_Z^{(dz)}\right)$,
\begin{equation}
E_{f}\left(f_Z^{(dz)}\right) \ge \max_{s\ge 0} \sum_{z_g}\phi(z_g, s)f_Z^{(dz)}(z_g) .
\label{ConditionalEf}
\end{equation}

{\bf Part III: } According to (\ref{EFCR1}), (\ref{SecondTerm}) and (\ref{ConditionalEf}), we have
\begin{eqnarray}
&& E_{Fc} \ge \min_{f_{Z}^{(dz)}, \sum_{z_g}z_gf_Z^{(dz)}(z_g) \in [1-dz, 1]} \left\{E_{f}(f_{Z}^{(dz)}) + \sum_{z_g} \frac{(z_g-1)^2}{2}f_Z^{(dz)}(z_g) \right \} -\frac{3}{2}dz \nonumber \\
&& \quad \ge \min_{f_{Z}^{(dz)}, \sum_{z_g}z_gf_Z^{(dz)}(z_g) \in [1-dz, 1]} \max_{s\ge 0} \sum_{z_g} \left(\phi(z_g, s)+\frac{(z_g-1)^2}{2}\right)f_Z^{(dz)}(z_g)  -\frac{3}{2}dz.
\label{EFCR2}
\end{eqnarray}

Define $E_{Fc}^{(0)}=\lim_{dz\to 0} E_{Fc} $. Let $f_Z$ be a probability density function defined for $z\in [0, \infty)$. Inequality (\ref{EFCR2}) implies that
\begin{eqnarray}
&& E_{Fc}^{(0)} \ge \min_{f_{Z}, \int_0^{\infty}zf_Z(z)dz=1} \max_{s\ge 0}  \int_{0}^{\infty} \left(\phi(z, s)+\frac{(z-1)^2}{2}\right)f_Z(z)dz \nonumber \\
&& \quad = \max_{s\ge 0} \min_{f_{Z}, \int_0^{\infty}zf_Z(z)dz=1}  \int_{0}^{\infty} \left(\phi(z, s)+\frac{(z-1)^2}{2}\right)f_Z(z)dz.
\label{EFCR3}
\end{eqnarray}

Assume that $f_Z^*$ is the density function minimizing the last term in (\ref{EFCR3}). If we can find $0<\lambda<1$, and two density functions $f_Z^{(1)}$, $f_Z^{(2)}$ with $\int_0^{\infty} zf_Z^{(1)}(z)dz =1$, $\int_0^{\infty} zf_Z^{(2)}(z)dz =1$, such that
\begin{equation}
f_Z^* = \lambda f_Z^{(1)}+ (1-\lambda)f_Z^{(2)},
\label{fZdecomposition}
\end{equation}
then it is easy to show that the last term in (\ref{EFCR3}) must be minimized either by $f_Z^{(1)}$ or $f_Z^{(2)}$. Since this contradicts the assumption that $f_Z^*$ is optimum, a nontrivial decomposition like (\ref{fZdecomposition}) must not be possible. Consequently, $f_Z^*$ can take non-zero values on at most two different $z$ values. Therefore, we can carry out the optimization in (\ref{EFCR3}) only over the following class of $f_Z$ functions, characterized by two variables $0\le z_0\le 1$ and $0 \le \gamma \le 1$,
\begin{equation}
f_Z(z)= \gamma \delta(z-z_0) + (1-\gamma) \delta\left(z- \frac{1-z_0\gamma}{1-\gamma} \right),
\label{OptEmpirical}
\end{equation}
where $\delta()$ is the impulse function.

Let us fix $\gamma$ first, and consider the following lower bound on $E_{Fc}^{(0)}(\gamma)$, which is obtained by substituting (\ref{OptEmpirical}) into (\ref{EFCR3}),
\begin{eqnarray}
E_{Fc}^{(0)}(\gamma)\ge \min_{0\le z_0\le 1}\max_{s\ge 0} \gamma\phi(z_0, s)+(1-\gamma)\phi\left(\frac{1-z_0\gamma}{1-\gamma}, s \right)+\frac{\gamma}{1-\gamma}\frac{(1-z_0)^2}{2}.
\label{EFCRpXrogamma}
\end{eqnarray}
Since given $z_0$, $\gamma\phi(z_0, s)+(1-\gamma)\phi\left(\frac{1-z_0\gamma}{1-\gamma}, s \right)$ is a linear function of $s$, depending on the value of $\gamma$, the optimum $s^*$ that maximizes the right hand side of (\ref{EFCRpXrogamma}) should satisfy either $s^*=E_z(z_0)$ or $s^*=E_z\left(\frac{1-z_0\gamma}{1-\gamma}\right)$.

When $\gamma \ge \frac{1-r_o}{2}$, we have $s^*=E_z(z_0)$. This yields
\begin{eqnarray}
E_{Fc}^{(0)} \ge \min_{0 \le z_0, \gamma \le 1} \left[ \frac{\gamma}{1-\gamma}\frac{(1-z_0)^2}{2} +  (1-r_o) E_z(z_0)  \right] .
\label{EFCRpXrogammalarge}
\end{eqnarray}
When $\gamma \le \frac{1-r_o}{2}$, we have $s^*=E_z\left(\frac{1-z_0\gamma}{1-\gamma}\right)$, which gives
\begin{eqnarray}
E_{Fc}^{(0)} &\ge & \min_{0 \le z_0, \gamma \le 1} \left[ 2\gamma E_z(z_0)+ \frac{\gamma}{1-\gamma}\frac{(1-z_0)^2}{2} + (1-r_o-2\gamma)E_z\left(\frac{1-\gamma z_0}{1-\gamma}\right)  \right].
\label{EFCRpXrogammasmall}
\end{eqnarray}
By substituting $E_z(z)=\max_{0\le \rho\le 1}[-\rho\frac{R}{r_o}+zE_0(\rho,p_X)]$ into (\ref{EFCRpXrogammasmall}), we get
\begin{eqnarray}\label{EFCRpXrogammasmal2}
&& E_{Fc}^{(0)} \ge \min_{0 \le z_0,\gamma \le 1} \max_{0\le \rho\le 1} \left\{ (1-r_o)\left[-\rho\frac{R}{r_o}+E_0(\rho,p_X)\right]- \right. \nonumber\\
&&\qquad\qquad\qquad \left.\frac{\gamma}{1-\gamma} \left[(1+r_o)(1-z_0)E_0(\rho,p_X)-\frac{(1-z_0)^2}{2}\right] \right\}.
\end{eqnarray}
Note that if $(1+r_o)(1-z_0)E_0(\rho,p_X)-\frac{(1-z_0)^2}{2}< 0$, then $E_{Fc}^{(0)}\ge(1-r_o)\left[-\rho\frac{R}{r_o}+E_0(\rho,p_X)\right]$ with the right hand side of the inequality equaling Forney's exponent for given $p_X$ and $r_o$. This contradicts with the fact that Forney's exponent is the maximum achievable exponent for one-level concatenated codes in a classical system \cite{ref Forney66}. Therefore, we must have $(1+r_o)(1-z_0)E_0(\rho,p_X)-\frac{(1-z_0)^2}{2}\ge 0$. Consequently, the right hand sides of both (\ref{EFCRpXrogammalarge}) and (\ref{EFCRpXrogammasmal2}) are minimized at the margin of $\gamma^*=\frac{1-r_o}{2}$. This gives
\begin{eqnarray}\label{EFCRpXrogammasmal3}
&& E_{Fc}^{(0)} \ge \min_{0 \le z_0 \le 1} \left\{(1-r_o)E_z(z_0) +\frac{1-r_o}{1+r_o}\frac{(1-z_0)^2}{2}\right\}\nonumber\\
&& = \min_{0 \le z_0 \le 1} \max_{0\le \rho\le 1} \left\{(1-r_o)\left(-\rho\frac{R}{r_o}+E_0(\rho,p_X)\right) +\frac{1-r_o}{1+r_o}\frac{(1-z_0)}{2}\left[ (1-z_0)-2(1+r_o)E_0(\rho,p_X)\right]\right\}. \quad
\end{eqnarray}

Note that if $\rho$ is chosen to satisfy $(1+r_o)E_0(\rho,p_X)\ge 1$, the last term in (\ref{EFCRpXrogammasmal3}) is minimized at $z_0^*=0$, which gives
\begin{equation}\label{EFCRpXrogammasmal4}
E_{Fc}^{(0)} \ge \max_{0\le \rho\le 1} \left\{-\rho\frac{R}{r_o}(1-r_o)+\frac{1-r_o}{1+r_o}\right\}.
\end{equation}
The right hand side of (\ref{EFCRpXrogammasmal4}) is maximized at $\rho^*=0$. However, $\rho=0$ implies $(1+r_o)E_0(\rho,p_X)=0<1$ which contradicts the assumption $(1+r_o)E_0(\rho,p_X)\ge 1$. Therefore, we can assume that $(1+r_o)E_0(\rho,p_X)\le 1$. Consequently, the last term in (\ref{EFCRpXrogammasmal3}) is minimized at $z_0^* =1-(1+r_o)E_0$. This gives
\begin{eqnarray}
&& E_{Fc}^{(0)} \ge \max_{0\le \rho\le 1} (1-r_o)\left(-\rho\frac{R}{r_o}+E_0(\rho,p_X)\left[1-\frac{1+r_o}{2}E_0(\rho,p_X)\right]\right).
\label{EFCPXro}
\end{eqnarray}

By optimizing (\ref{EFCPXro}) over $p_X$ and $r_o$, it can be seen that the error exponent given in (\ref{EFC}) is achievable if we first take $N_o$ to infinity and then take $N_i$ to infinity.

{\bf Part IV: }
To achieve linear coding complexity, let us assume that $N_i$ is fixed at a large constant while $N_o$ is taken to infinity. According to \cite{ref Guruswami05}, it is easy to see that the encoding complexity is linear in the number of transmitted symbols\footnote{In other words, we assume that no encoding complexity is spent on codeword symbols that are not transmitted.}. At the receiver, we keep at most $2N_i$ symbols for each inner code and drop the extra received symbols. Consequently, the effective codeword length parameter of any inner code is upper-bounded by $2$. Because (\ref{EFCRpXrogammalarge}) and (\ref{EFCRpXrogammasmal2}) are both minimized at $\gamma^*=\frac{1-r_o}{2}$, according to (\ref{OptEmpirical}), the empirical density function $f_Z(z)$ that minimizes the error exponent bound takes the form $f_Z(z)= \frac{1-r_o}{2} \delta(z-z_0) + \frac{1+r_o}{2} \delta\left(z- \frac{2-z_0(1-r_o)}{1+r_o} \right)$, with $z_0, \frac{2-z_0(1-r_o)}{1+r_o} <2$. Therefore, upper bounding the effective codeword length parameter by $2$ does not change the error exponent result. However, with $z_k\le 2$, $\forall k$, the decoding complexity of any inner code is upper-bounded by a constant in the order of $O(\exp(2N_i))$. According to \cite{ref Zheng09}, the overall decoding complexity of the concatenated code is therefore linear in $N_o$, and hence is linear in $N$. Since fixing $N_i$ causes a reduction of $\zeta_1>0$ in the achievable error exponent, and $\zeta_1$ can be made arbitrarily small as we increase $N_i$, we conclude that fountain error exponent $E_{Fc}(R)$ given in (\ref{EFC}) can be {\it arbitrarily approached} by one-level concatenated fountain codes with a linear coding complexity.
\end{proof}

\subsection{Proof of Corollary \ref{Corollary1}}
\label{ProofCorollary1}
\begin{proof}
Because $0\le r_o \le 1$, it is easy to see $\tilde{E}_{Fc}(R)\le E_{Fc}(R) \le E_c(R)$. We will next prove $\lim_{R\rightarrow \mathcal{C}_F}\frac{\tilde{E}_{F_c}(R)}{E_{F_c}(R)}=1$.

Define $g(p_X,r_o,\rho) = (1-r_o)\left(-\rho\frac{R}{r_o} + E_0(\rho,p_X)\left[1-\frac{1+r_o}{2}E_0(\rho,p_X)\right]\right)$, such that 
\begin{equation}
E_{F_c}(R) = \max_{p_X, \frac{R}{\mathcal{C}_F}\le r_o\le1,0 \le \rho \le 1}g(p_X,r_o,\rho).
\end{equation}
Using Taylor's expansion to expand $g(p_X,r_o,\rho)$ at $r_o = 1$ and $\rho = 0$, we get
\begin{equation}
g(p_X,r_o,\rho) = \sum_{i,j}\frac{1}{(i+j)!} \beta(i,j) (r_o-1)^i \rho ^j  ,
\end{equation}
where $\beta(i,j)=\left.\frac{\partial^{(i+j)} g(p_X,r_o,\rho)}{\partial r_o^i \partial \rho^j}\right|_{r_o = 1,\rho = 0}$, with $i$ and $j$ being nonnegative integers. It can be verified that $\beta(i,j)=0$ if $i=0$ or $j=0$.
%
%\begin{eqnarray}
%&&\beta(1,0) = \left.\left\{\rho\frac{R}{r_o^2}-E_0(\rho,p_X)+r_oE^2_0(\rho,p_X)\right\}\right|_{r_o=1,\rho=0} = 0,\nonumber\\
%&&\beta(2,0) = \left.\left\{-\rho\frac{2R}{r_o^3}+E^2_0(\rho,p_X)\right\}\right|_{r_o=1,\rho=0} = 0,\nonumber\\
%&&\beta(i,0) = \left.\left\{-\rho\frac{i! (-1)^i R}{r_o^{i+1}}\right\}\right|_{r_o=1,\rho=0} = 0, \qquad \forall i\ge 3.
%\label{betai0}
%\end{eqnarray}
%It can also be verified that
%\begin{eqnarray}
%&&\beta(0,1) = \left. (1-r_o)\left(-\frac{R}{r_o}+\frac{\partial E_0(\rho,p_X)}{\partial \rho}-\frac{\partial E_0(\rho,p_X)}{\partial \rho}E_0(\rho,p_X)(1+r_o)\right)\right|_{r_o = 1,\rho = 0} =0, \nonumber\\
%&&\beta(0,j) = \left. (1-r_o)h_j(R, \rho, r_o)\right|_{r_o = 1,\rho = 0} =0, \qquad \forall j\ge 2,
%\label{beta0j}
%\end{eqnarray}
%where $h_j(R, \rho, r_o)$ is a function of $R, \rho, r_o$.
We also have
\begin{eqnarray}
&&\beta(1,1) = \left.\left\{\frac{R}{r_o^2}-\frac{\partial E_0(\rho,p_X)}{\partial \rho}+2r_o E_0(\rho,p_X)\frac{\partial E_0(\rho,p_X)}{\partial \rho}\right\}\right|_{r_o=1,\rho=0}=R-\mathcal{C}_F, \nonumber\\
&&\beta(2,1) = -2R\neq 0, \nonumber \\
&& \beta(1,2) = \left.-\left\{\frac{\partial^2E_0(\rho,p_X)}{\partial \rho^2}-2\left(\frac{\partial E_0(\rho,p_X)}{\partial \rho}\right)^2\right\}\right|_{\rho=0} \ne 0.
\label{betaij}
\end{eqnarray}

Similarly, define $\tilde{g}(p_X,r_o,\rho) = (1-r_o)\left(-\rho\frac{R}{r_o} + E_0(\rho,p_X)\left[1-E_0(\rho,p_X)\right]\right)$, such that 
\begin{equation}
\tilde{E}_{F_c}(R) = \max_{p_X, \frac{R}{\mathcal{C}_F}\le r_o\le1,0 \le \rho \le 1}\tilde{g}(p_X,r_o,\rho).
\end{equation}
Using Taylor's expansion to expand $\tilde{g}(p_X,r_o,\rho)$ at $r_o = 1$ and $\rho = 0$, we get
\begin{eqnarray}
&& \tilde{g}(p_X,r_o,\rho) = \sum_{i,j}\frac{1}{(i+j)!} \tilde{\beta}(i,j) (r_o-1)^i \rho ^j.
\end{eqnarray}
where $\tilde{\beta}(i,j) = \left.\frac{\partial^{(i+j)} \tilde{g}(p_X,r_o,\rho)}{\partial r_o^i \partial \rho^j}\right|_{r_o = 1,\rho = 0}$.
%It can be verified that
%\begin{eqnarray}
%&&\tilde{\beta}(1,0) = \left.\left\{\rho\frac{R}{r_o^2}-E_0(\rho,p_X)+E^2_0(\rho,p_X)\right\}\right|_{r_o=1,\rho=0} = 0,\nonumber\\
%&&\tilde{\beta}(i,0) = \left.\left\{-\rho\frac{i!(-1)^iR}{r_o^{i+1}}\right\}\right|_{r_o=1,\rho=0} = 0, \qquad \forall i\ge 2.
%\label{tildebetai0}
%\end{eqnarray}
Similarly, we have $\tilde{\beta}(i,j)=0$ if $i=0$ or $j=0$ and $\tilde{\beta}(1,1)=\beta(1,1)=R-\mathcal{C}_F$, $\tilde{\beta}(2,1)=\beta(2,1)\ne 0$, $\tilde{\beta}(1,2)=\beta(1,2)\ne 0$.

By L'Hospital's rule, the following equality holds,
\begin{equation}
 \lim_{R\rightarrow \mathcal{C}_F}\frac{\tilde{E}_{Fc}(R)}{E_{Fc}(R)}=\lim_{R\rightarrow \mathcal{C}_F, r_o\rightarrow 1, \rho \rightarrow 0 } \frac{\frac{1}{2} \tilde{\beta}(1,1) (r_o-1) \rho + \frac{1}{6} \tilde{\beta}(2,1) (r_o-1)^2 \rho + \frac{1}{6} \tilde{\beta}(1,2) (r_o-1) \rho^2}{\frac{1}{2} \beta(1,1) (r_o-1) \rho + \frac{1}{6} \beta(2,1) (r_o-1)^2 \rho + \frac{1}{6} \beta(1,2) (r_o-1) \rho^2}=1.
\end{equation}
\end{proof}

%It can also be verified that
%\begin{eqnarray}
%&&\tilde{\beta}(0,1) = \left. (1-r_o)\left(-\frac{R}{r_o}+\frac{\partial E_0(\rho,p_X)}{\partial \rho}-\frac{\partial E_0(\rho,p_X)}{\partial \rho}2E_0(\rho,p_X)\right)\right|_{r_o = 1,\rho %= %0} = 0.\nonumber\\
%&&\tilde{\beta}(0,j)= \left. (1-r_o)\tilde{h}_j(R, \rho, r_o)\right|_{r_o = 1,\rho = 0} =0, \qquad \forall j\ge 2,
%\label{tildebeta0j}
%\end{eqnarray}
%where $\tilde{h}_j(R, \rho, r_o)$ is a function of $R, \rho, r_o$. We also have
%\begin{eqnarray}
%&&\tilde{\beta}(1,1) = \left. \left\{\frac{R}{r_o^2}-\frac{\partial E_0(\rho,p_X)}{\partial \rho}+2E_0(\rho,p_X)\frac{\partial E_0(\rho,p_X)}{\partial \rho}\right\}\right|_{r_o=1,\rho=0} = %R-C_F,\nonumber\\
%&&\tilde{\beta}(2,1) = -2R \neq 0, \nonumber \\
%&& \tilde{\beta}(1,2)=\left.-\left\{\frac{\partial^2E_0(\rho,p_X)}{\partial \rho^2}-2\left(\frac{\partial E_0(\rho,p_X)}{\partial \rho}\right)^2\right\}\right|_{\rho=0} \neq  0,
%\label{tildebetaij}
%\end{eqnarray}

%Because $\beta(1,1)=\tilde{\beta}(1,1)$, $\beta(2,1)=\tilde{\beta}(2,1)\ne 0$, $\beta(1,2)=\tilde{\beta}(1,2)\ne 0$, by L'Hospital's rule, we have
%\begin{eqnarray}
%&& \lim_{R\rightarrow \mathcal{C}_F}\frac{\tilde{E}_{Fc}(R)}{E_{Fc}(R)}=\lim_{R\rightarrow \mathcal{C}_F, r_o\rightarrow 1, \rho \rightarrow 0 } \frac{\frac{1}{2} \tilde{\beta}(1,1) (r_o-1) %\rho + \frac{1}{6} \tilde{\beta}(2,1) (r_o-1)^2 \rho + \frac{1}{6} \tilde{\beta}(1,2) (r_o-1) \rho^2}{\frac{1}{2} \beta(1,1) (r_o-1) \rho + \frac{1}{6} \beta(2,1) (r_o-1)^2 \rho + \frac{1}{6} %\beta(1,2) (r_o-1) \rho^2}=1. \nonumber \\
%\end{eqnarray}
%\end{proof}

\subsection{Proof of Theorem \ref{Theorem4}}
\label{ProofTheorem4}
\begin{proof}
Define
\begin{eqnarray}
&& \hat{g}(\gamma, r_o, \rho) = (1-r_o) \left(\rho I(p_X) \left(1-\frac{\gamma}{r_o}\right) +\frac{\rho^2}{2}\left(\left. \frac{\partial^2 E_0(\rho, p_X)}{\partial \rho^2}\right|_{\rho=0}-2I^2(p_X)\right)\right), \nonumber \\
&& \hat{E}_{Fc}(\gamma, p_X, r_o) = \max_{0\le\rho\le 1}\hat{g}(\gamma, r_o, \rho), \nonumber \\
&& \hat{E}_{Fc}(\gamma, p_X) = \max_{0\le r_o \le 1} \hat{E}_{Fc}(\gamma, p_X, r_o).
\end{eqnarray}
We will first prove that
\begin{equation}
\lim_{\gamma \rightarrow 1} \frac{E_{Fcs}(\gamma, p_X)}{\hat{E}_{Fc}(\gamma, p_X)}=1.
\end{equation}

Note that $\hat{g}(\gamma, r_o, \rho)$ is maximized at $\rho^*=\frac{I(p_X) \left(1-\frac{\gamma}{r_o}\right) }{\left. -\frac{\partial^2 E_0(\rho, p_X)}{\partial \rho^2}\right|_{\rho=0}+2I^2(p_X)}$, where we have assumed that $0\le \rho^* \le 1$. This assumption is valid when $r_o$ is also optimized. Consequently, $\hat{E}_{Fc}(\gamma, p_X, r_o)$ is maximized at $r_o^*=\mathop{\arg\max}_{0\le r_o \le 1} (1-r_o)\left(1-\frac{\gamma}{r_o}\right)^2=\frac{\sqrt{\gamma^2+8\gamma}-\gamma}{2}$. Therefore,
\begin{equation}
\lim_{\gamma \rightarrow 1} \frac{E_{Fcs}(\gamma, p_X)}{\hat{E}_{Fc}(\gamma, p_X)}\ge \lim_{\gamma \rightarrow 1} \left.\left[\frac{E_{Fcs}(\gamma, p_X, \rho)}{\hat{g}(\gamma, p_X, \rho, r_o)}\right|_{ \rho =\rho^*, r_o=r_o^*} \right]= 1.
\label{Limit1}
\end{equation}

Following a similar idea as the proof of Corollary \ref{Corollary1}, it can be shown that
\begin{equation}
\lim_{\gamma \rightarrow 1} \frac{\hat{E}_{Fc}(\gamma, p_X)}{E_{Fc}(\gamma, p_X)}=1.
\label{Limit2}
\end{equation}
Combining (\ref{Limit1}) and (\ref{Limit2}), we get
\begin{equation}
\lim_{\gamma \rightarrow 1} \frac{E_{Fcs}(\gamma, p_X)}{E_{Fc}(\gamma, p_X)}=\lim_{\gamma \rightarrow 1} \frac{E_{Fcs}(\gamma, p_X)}{\hat{E}_{Fc}(\gamma, p_X)}
\lim_{\gamma \rightarrow 1} \frac{\hat{E}_{Fc}(\gamma, p_X)}{E_{Fc}(\gamma, p_X)} \ge 1.
\label{Limit3}
\end{equation}
Because $E_{Fcs}(\gamma, p_X) \le E_{Fc}(\gamma, p_X)$, (\ref{Limit3}) implies $\lim_{\gamma \rightarrow 1} \frac{E_{Fcs}(\gamma, p_X)}{E_{Fc}(\gamma, p_X)}=1$.

\end{proof}

%\section*{Acknowledgment}

%\section*{Acknowledgment}
% optional entry into table of contents (if used)
%\addcontentsline{toc}{section}{Acknowledgment}



% http://www.ctan.org/tex-archive/macros/latex/contrib/supported/IEEEtran/bibtex
%\bibliographystyle{IEEEtran.bst}
% (used to reserve space for the reference number labels Diamondsuit)

%all figures are of width 3.5 inch

\newpage 

\begin{figure}[h]
  \centering
  \includegraphics[width=5 in]{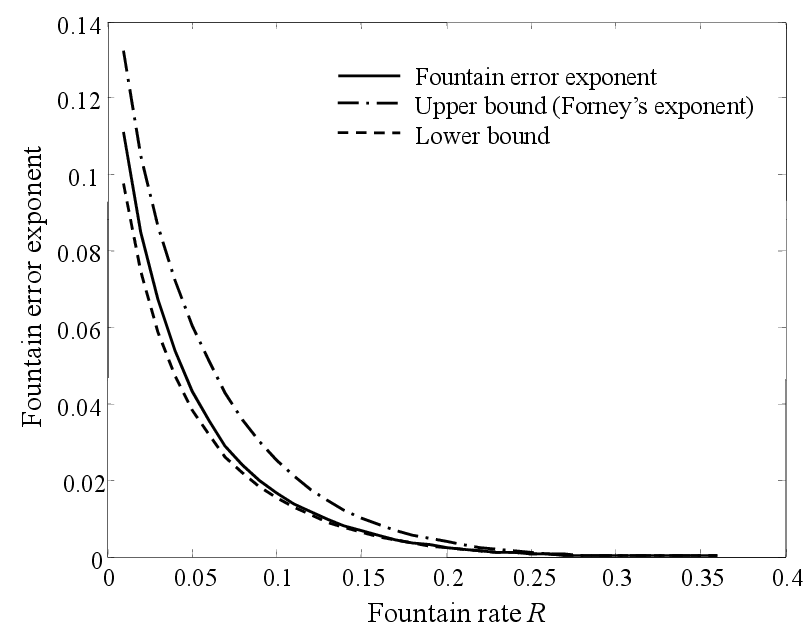}\caption{\label{Fig3} Comparison of fountain error exponent $E_{Fc}(R)$, its upper bound $E_c(R)$, and its lower bound $\tilde{E}_{Fc}(R)$.}
\end{figure}

\begin{figure}[h]
  \centering
  \includegraphics[width=5 in]{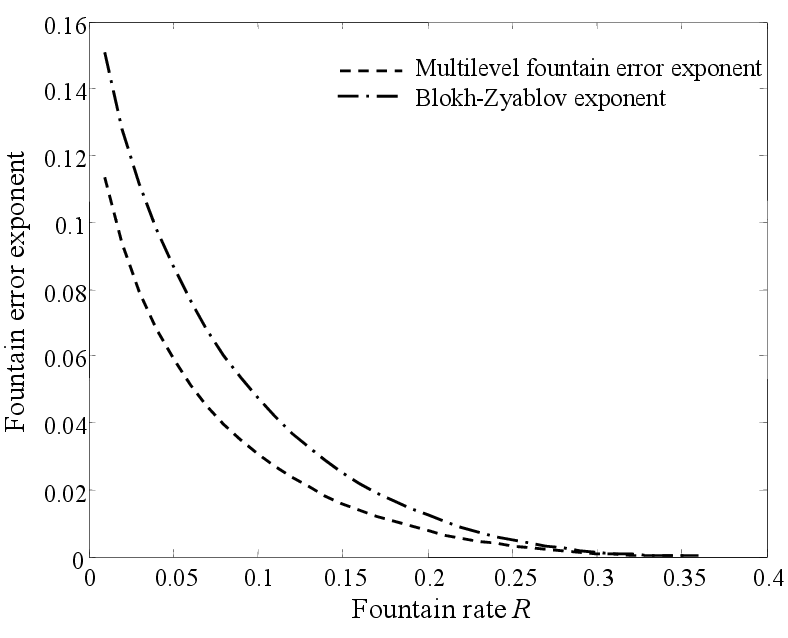}\caption{\label{Fig4} Comparison of mulit-level fountain error exponent $E_{Fc}^{(\infty)}(R)$ and the Blokh-Zyablov exponent $E_c^{(\infty)}(R)$.}
\end{figure}

\begin{figure}[h]
  \centering
  \includegraphics[width=5 in]{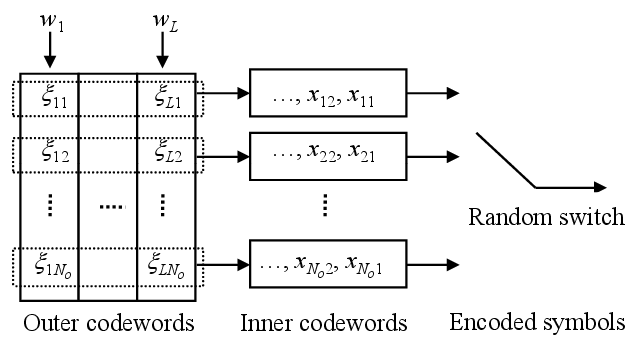}
  \caption{\label{Fig5} Concatenated fountain codes for rate compatible communication.}
\end{figure}

\begin{figure}[h]
  \centering
  \includegraphics[width=5 in]{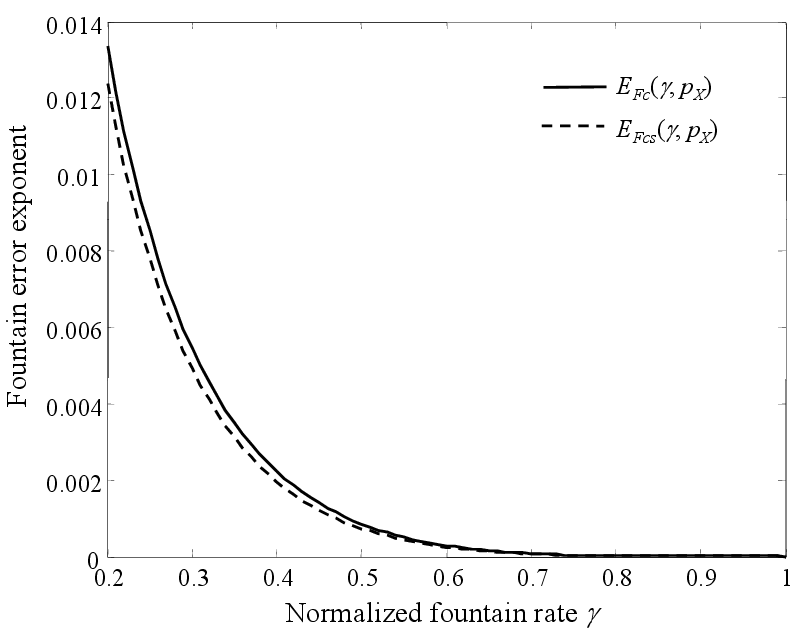}
  \caption{\label{Fig6} Error exponents achieved by optimal $r_o$ and suboptimal $r_o=\frac{\sqrt{\gamma^2+8\gamma}-\gamma}{2}$ versus normalized fountain rate $\gamma$.}
\end{figure}


%\vfill


\end{document}